\documentclass[twocolumn,superscriptaddress,showpacs,aip]{revtex4}
\usepackage{mathrsfs}
\usepackage{amssymb, amsbsy, amsmath, latexsym, dsfont, array, layout, graphicx,mathrsfs,color}

\newcommand{\one}{\mathds{1}}
\newcommand{\ket}[1]{\left|{#1}\right\rangle}
\newcommand{\bra}[1]{\left\langle{#1}\right|}

\begin{document}

\title{Trapping photons on the line: controllable dynamics of a quantum walk}
\author{P. Xue\footnote{gnep.eux@gmail.com}
} \affiliation{Department of Physics, Southeast University, Nanjing
211189, China} \affiliation{State Key Laboratory of Precision
Spectroscopy, East China Normal University, Shanghai 200062, China}
\author{H. Qin}
\affiliation{Department of Physics, Southeast University, Nanjing
211189, China}
\author{B. Tang}
\affiliation{Department of Physics, Southeast University, Nanjing
211189, China}

\begin{abstract}
We demonstrate a coined quantum walk over ten steps in a
one-dimensional network of linear optical elements. By applying
single-point phase defects, the translational symmetry of an ideal
standard quantum walk is broken resulting in localization effect in
a quantum walk architecture. We furthermore investigate how the
level of phase due to single-point phase defects and coin settings
influence the strength of the localization signature.
\end{abstract}

\maketitle

Quantum walks (QWs)~\cite{ADZ93} are the quantum mechanical analog
of classical random walks (RWs), and hence can be used to develop
quantum algorithms~\cite{Ambainis,Spielman,Whaley02,Kempe03}, emerge
as an alternative to the standard circuit model for quantum
computing~\cite{Childs09,Childs13,Lovett10}, and represent one of
the most promising resources for the simulation of physical system
and important phenomena such as topological phases~\cite{KRBE10},
energy transport in photosynthesis~\cite{OPR06,HSW10}, Anderson
localization~\cite{A58,BCA03,W12,ES11,YKE08,C13,SS11} and quantum
chaos~\cite{WM04,BB04,B06,Xue13}.

A standard model of a one-dimension (1D) discrete-time QW consists
of a quantum walker carrying a quantum coin. The walker goes back
and forth along a line and the direction at each step depends on the
result of a coin flip, which can be implemented by an arbitrary
unitary operation in SU(2) following by a conditional position shift
operation. The position variance of the walker $\sigma^2=\langle
x^{2}\rangle-\langle x\rangle^2$ is linear on the number of the
steps for RWs and quadratic for QWs. The position distribution of a
standard QW $P(x)$ shows a ballistic diffusion and that of RWs
diffusive spreading. Furthermore if the static disorder is
introduced in the dynamics of QWs, by changing the interference
pattern localization effect can be observed in the QW
architecture---spreading more slowly than RWs~\cite{W12}.

Experimental QWs began in 1999 and were performed with the frequency
space of an optical resonator~\cite{B99}. Several alternative
realizations were quickly afterwards based on energy levels in
nuclear magnetic resonance~\cite{DH03}, phase and position space of
trapped ions~\cite{ZR10,SM09,KF09} and trapped neutral
atoms~\cite{CG06}, photons in beam splitter array, in fiber loop and
in waveguide
structures~\cite{Xue13,D05,Z07,PO10,PS08,SS12,BF+10,S10,SSV+12}.

\begin{figure}
\includegraphics[width=8.5cm]{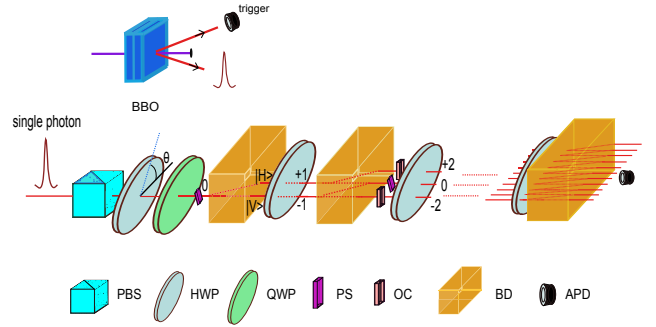}
\caption{(Color online.) Detailed sketch of the setup for $10$-step
QW with SPPDs. Single-photons created via type-I SPDC are injected
to the optical network. Arbitrary initial coin states are prepared
by a PBS, HWP and QWP. PSs are placed in the corresponding spatial
modes and the optical compensators (OCs) are used to compensate the
temporal delay caused by PSs. Coincident detection of photons at
APDs ($7$ns time window) predicts a successful run of the
QW.}\label{Fig1}
\end{figure}

In this work, we report on the implementation of a discrete QW with
single-point phase defects (SPPDs), and implement methods suggested
in~\cite{W12}. We investigate the evolution of single-photons moving
in a discrete environment presenting SPPDs, casting light on the
natural feature of the physical description of QWs such as general
properties of diffusion modified by quantum or interference effects.
Compared to the previous optical experiments on standard QWs (e.g.
without static disorder)~\cite{PO10,PS08,SS12,BF+10,S10,SSV+12}, we
use phase shifters (PSs) for realization of the position-dependent
phases acquired by the walker at the certain position. Compared to
the experimental realization of QWs with time-dependent phase
defects~\cite{SS11}, our experiment on QWs with position-dependent
phases is more close to nature and can be used to investigate
localization effect on low-dimensional structure, which would be
interesting in research on properties of low-dimensional materials.

We use the beam-displacer array as interferometer network similar to
the setup in~\cite{Xue13}. By taking advantage of the intrinsically
stable interferometers, our approach is robust and able to control
both coin and walker at each step. Benefiting from the fully
controllable implementation, we experimentally study the impact of
the SPPD and coin bias on the localization effect in a QWs
architecture and the experimental results agree with the theoretical
predictions. Compared to the previous experimental results which
only simulated localization effect by trapping the walker in the
original position $x=0$, we experimentally localize the
single-photons in different positions.

In our experiment, we are able to achieve $10$ QW-steps. 
The challenge of our experiment is to realize specific
polarizing-independent phase on each site via microscope slides with
precise effective thickness as PSs and to keep high interference
visibility for each step even with phase defect. By introducing
controllable PSs in paths of the interferometers, We have managed to
create these versatile interferometer networks which can be used in
many other fields.

The non-degenerated polarization degenerate photon pairs generated
via type-I spontaneous parametric downconversion in two
$0.5$mm-thick nonlinear-$\beta$-barium-borate (BBO) crystals cut at
$29.41^o$, pumped by a $400.8$nm CW diode laser (LBX-405-100,
Oxxius) with up to $100$mW of power. For 1D QWs, by triggering on
one photon, the other at wavelength $800$nm is prepared into a
single-photon state. A polarizing beam splitter (PBS) following by
waveplates allow generation of any polarized state of single photon
(e.g. any initial coin state). Interference filters determine the
photon bandwidth $3$nm and then individual downconverted photons are
steered into the optical modes of the linear-optical network formed
by a series of birefringent calcite beam displacers (BDs), half-wave
plates (HWPs) and PSs. Output photons are detected using avalanche
photo-diodes (APDs, SPCM-AQRH-14-FC) with dark count rate of
$<100$s$^{-1}$ whose coincident signals---monitored using a
commercially available counting logic (id800-TDC)---are used to
postselect two single-photon events. The total coincident counts are
about $300$s$^{-1}$ (the coincident counts are collected over
$60$s). The probability of creating more than one photon pair is
less than $10^{-4}$ and can be neglected.

The coin state is encoded in the polarization $\ket{H}$ and
$\ket{V}$ of the input photon. In the basis $\{\ket{H},\ket{V}\}$,
the coin operation for each step is given by
\begin{equation}
C(\theta)=\begin{pmatrix}\cos2\theta & \sin2\theta\\
        \sin2\theta & -\cos2\theta\end{pmatrix}
\end{equation}
        with $\theta\in\left(0^o,45^o\right)$, consisting of a
polarization rotation, which is realized with a HWP setting. For
example, the Hadamard operator is realized with a HWP set to
$\theta=22.5^o$.

\begin{figure*}
\includegraphics[width=4.2cm]{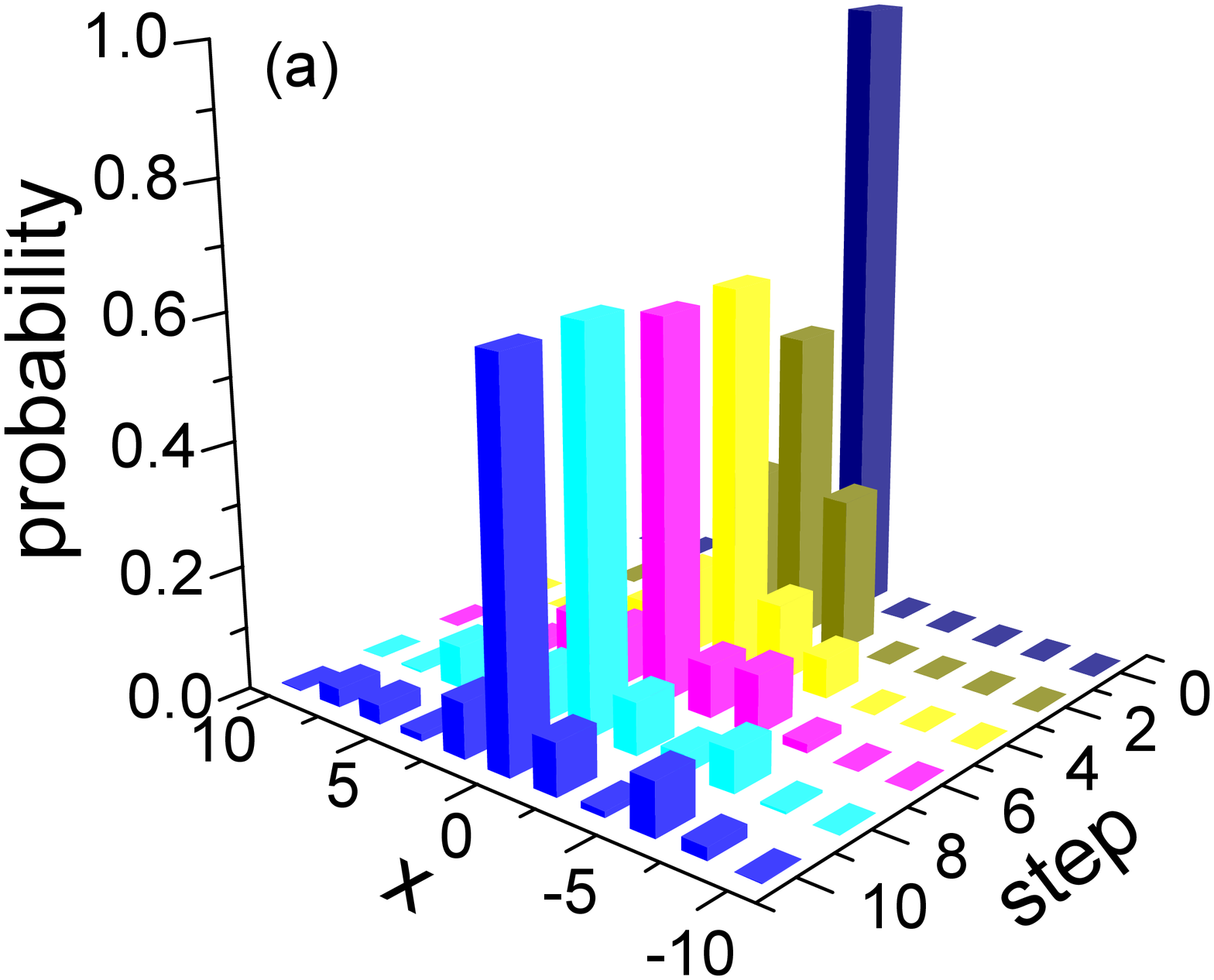}
\includegraphics[width=4.2cm]{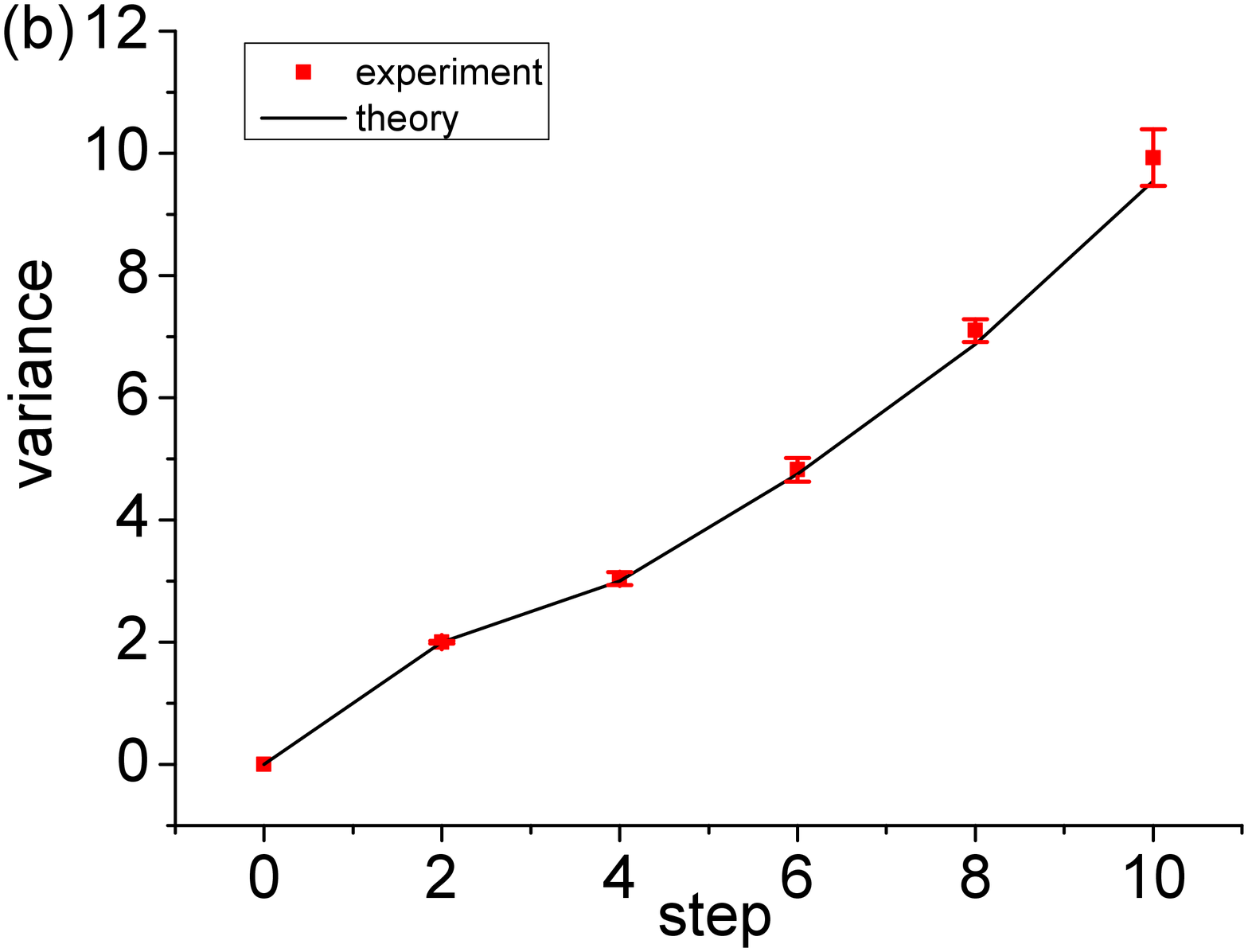}
\includegraphics[width=4.2cm]{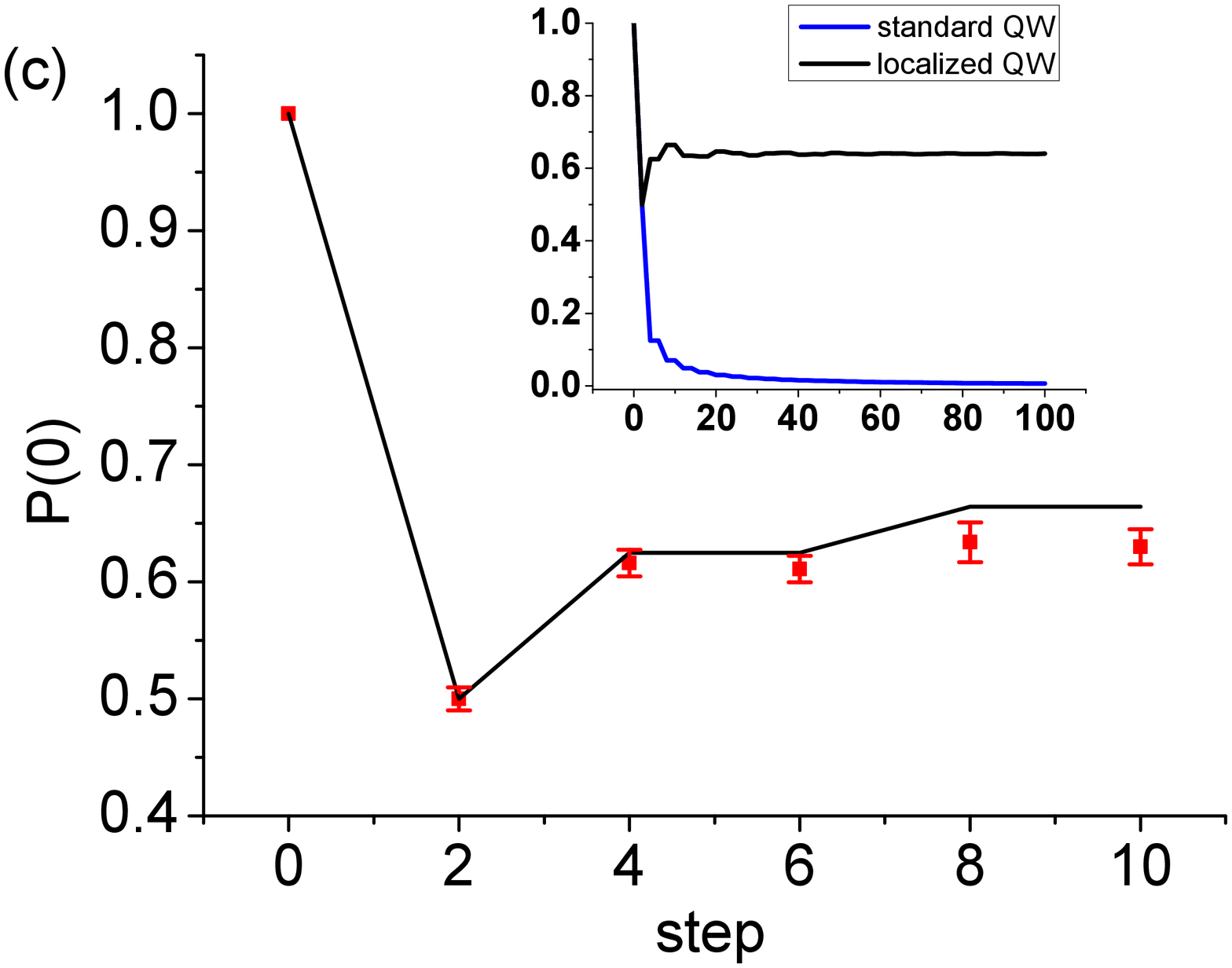}
\caption{(Color online.) (a) Experimental position distributions for
successive steps of the Hadamard ($C(22.5^o)$) QW with the SPPD
$\phi=180^o$ in the original position $x=0$ and antisymmetric
initial coin state up to $10$ steps. (b) Measured position variance
and (c) recurrence probability of localized QW for $1$ to $10$
steps, compared to theoretical predictions (solid lines). Error bars
indicate the statistical uncertainty. Inset shows comparison of
theoretical prediction of the recurrence probability of localized
Hadamard QW with the SPPD $\phi=180^o$ and standard QW with
$\phi=0$.}\label{Fig2}
\end{figure*}

The walker's positions are represented by longitudinal spatial
modes. The unitary operator for each step of a QW with a SPPD
\begin{equation}
U=\left(S_{\phi_n}\otimes\ket{H}\bra{H}+S_{\phi_n}^\dagger\otimes\ket{V}\bra{V}\right)\left(\one\otimes
C(\theta)\right)
\end{equation} with the position shift operator $S_{\phi_n}\ket{x}=e^{i\phi\delta(x-n)}\ket{x-1}$ ($S_{\phi_n}^\dagger
\ket{x}=e^{i\phi\delta(x-n)}\ket{x+1}$) on the modes manipulates the
wavepacket to propagate according to the polarization of the
photons. The translational symmetry of an ideal standard QW without
SPPD is now broken by modifying the phase of the walker on each
site, which can be realized by simply introducing PSs in the
specific interferometer arms. PSs are placed in the certain modes
$x=n$. By adjusting the relative angle between the PS and the
following BD the effective thickness of the PS changes and the
specific phase $\phi$ can be realized.

The spatial mode is implemented by a birefringent calcite BD with
length $28.165$mm and clear aperture $33$mm$\times 15$mm. The
optical axis of each BD is cut so that vertically polarized light is
directly transmitted and horizontal light undergoes a $3$mm lateral
displacement into a neighboring mode which interferes with the
vertical light in the same mode. Each pair of BDs forms an
interferometer. Only odd (even) sites of the walker are labeled at
each odd (even) step, since the probabilities of the walker
appearing on the other sites are zero.

The first $10$ steps of the QW with SPPD $\phi$ applied in the
original position $x=0$ are shown in Fig.~1 in detailed. The
longitudinal spatial modes after the $1$st step are recombined
interferometrically at the $2$nd step. The interference visibility
is reached $0.998$ per step (extinction ratio $1000:1$). 
The probabilities are obtained by normalizing photon counts on each
site to total number of photon counts for the respective step. The
measured probability distributions for $1$ to $10$ steps of a
Hadamard QW with SPPD $\phi=180^o$ and the antisymmetric initial
coin state $\left(\ket{H}-i\ket{V}\right)/\sqrt{2}$ are shown in
Fig.~2a. An average distance
$d=\frac{1}{2}\sum_x\left|P^\text{exp}(x)-P^\text{th}(x)\right|$ is
$0.046$ ensuring a good agreement between the measured probabilities
and theoretic predictions after $10$ steps. The walker state after
$4$ steps clearly shows the characteristic shape of a localization
distribution: a pronounced peak of the probability $0.615\pm0.011$
in the original position $x=0$ and the low probabilities in the side
positions. In contrast to the ideal standard Hadamard QW the
expansion of the wavepacket is highly suppressed and the probability
of the walker returning to the original position is enhanced
strictly and displays the signature of the localization effect. In
Figs.~2b and 2c, the position variance and recurrence probability,
i.e., the probability of the walker returning to the original
position, show the spread of the localized QW is much slower than
that of the standard QW and the walker is trapped in the original
position with high probability after the $4$th step. While in the
case of the $10$-step standard Hadamard QW without SPPD the variance
is given by $\sigma^2_Q=29.951$, a lower variance occurs in the RW
case with $\sigma^2_R=10$. Our measured value
$\sigma^2=9.586\pm0.463$ agrees well with the theoretical prediction
$9.547$ and shows an even slower spread than RW. The presented error
bars include only statistical errors, calculated from the standard
deviations of the values calculated by the Monte Carlo method. The
measured recurrence probability exhibits the localization effect of
QW with SPPD after the $4$th step. Compared to the walker in a
standard QW with SPPD showing a ballistic behavior, it is always
trapped in the position $x=0$ with high probability about $0.64$
after $4$ steps shown in Fig.~2c.

\begin{figure}
\includegraphics[width=4.5cm]{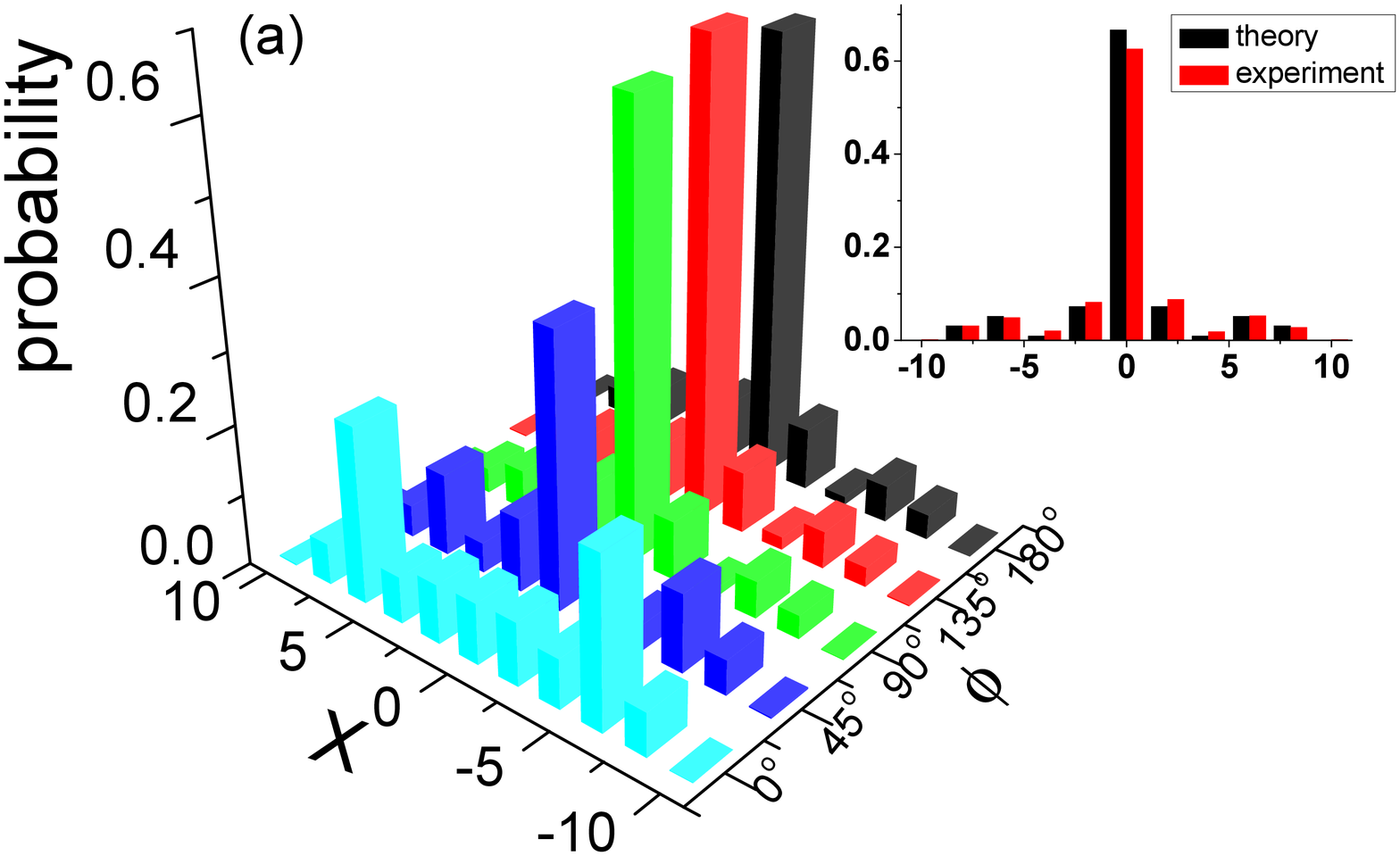}
\includegraphics[width=4.cm]{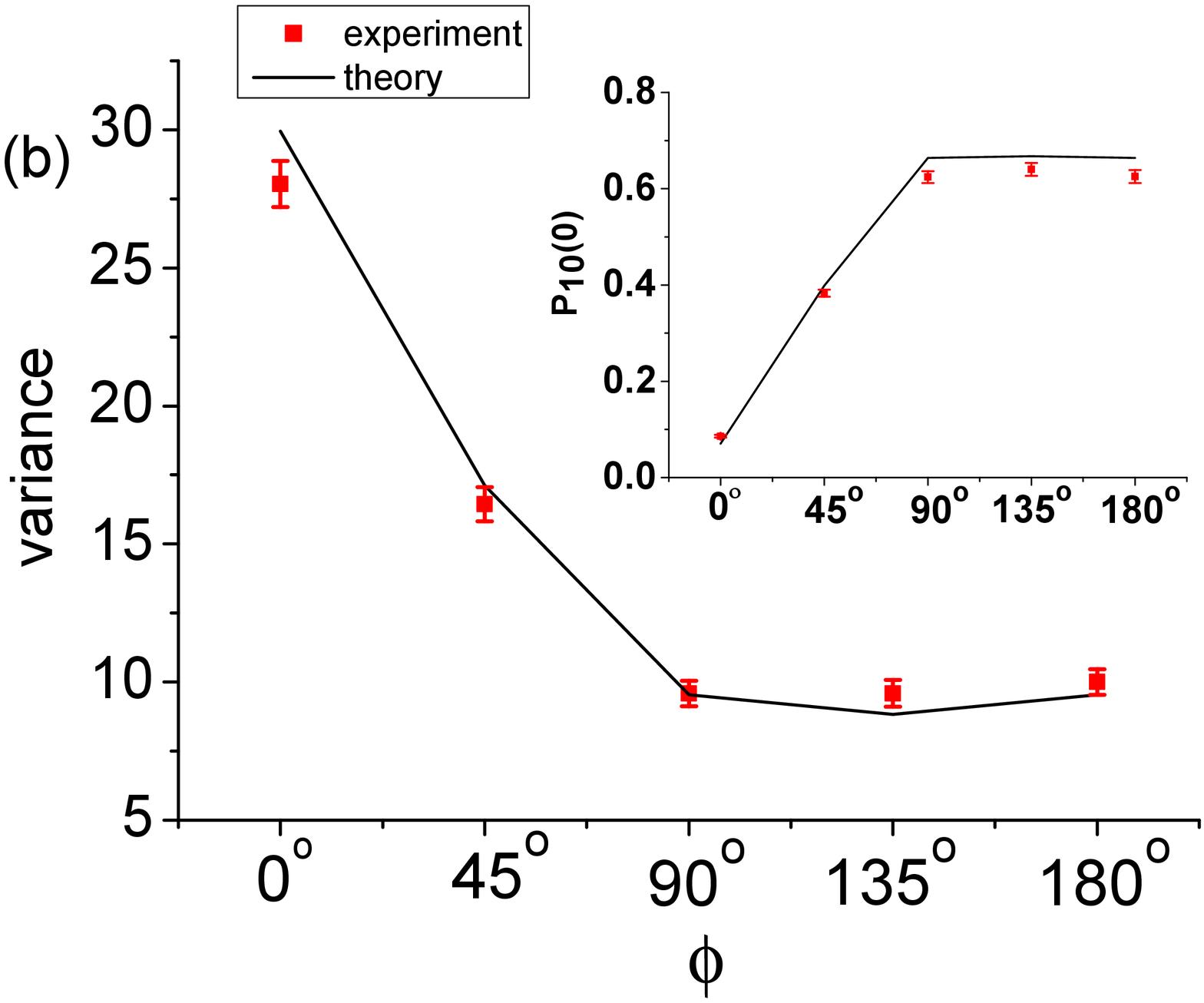}
\caption{(Color online.) (a) Experimental data of probability
distributions of the $10$-step Hadamard QW with antisymmetric
initial coin state and various single-point phase $\phi$ in the
original position $x=0$. The inset shows the probability
distribution of the $10$th-step Hadamard QW with SPPD $\phi=135^o$.
Red and black bars show experimental data and theoretical
predictions respectively. (b) Measured position variance of the
$10$-step Hadamard QW v.s. $\phi$, compared to theoretical
predictions (solid lines). The inset shows the measured recurrence
probabilities $P_{10}(0)$ after the $10$ steps as a function of
$\phi$.}\label{Fig3}
\end{figure}

Our second experimental result highlights the full control of the
implementation of the QW. In Fig.~3, we show the impact of
single-point phase $\phi\in\left[0^o,180^o\right]$ on the
localization effect. In this case, we change the effective thickness
of the PSs to realize different SPPDs in the original position.
Fig.~3a shows the position distribution of the $10$-step Hadamard QW
changes as a function of the single-point phase $\phi$. At the
$10$th step the recurrence probability $P_{10}(0)$ and the position
variance $\sigma^2$ as functions of $\phi$ are shown in Fig.~3b. For
the antisymmetric initial coin state, the recurrence probability
$P_{10}(0)$ increases with $\phi$ in the range
$\left[0^o,135^o\right]$ and decreases in the range
$\left(135^o,180^o\right]$. The maximal probability is achieved
(measured as $0.660\pm0.012$ and numerical simulated as $0.667$)
with $\phi=135^o$. The localization effect occurs in the range
$\left[45^o,180^o\right]$, which agrees with the analytic result.
Thus with fixed coin toss and initial state, whether or not the
localization effect can be observed depends on the choices of
single-point phase applied in the original position.

The dependence of the localization effect on the phase $\phi$ can be
explained~\cite{W12} by the overlap between the localized
eigenstates of the unitary operation $U$ and the initial state of
the walker+coin system shown in Fig.~4a. The number of the localized
eigenstates of $U$ depends on $\phi$. In the range
$\phi\in\left(0^o,45^o\right)$ and $\phi\in\left(135^o,180^o\right)$
there are two localized eigenstates. Whereas, in the range
$\phi\in\left(45^o,135^o\right)$ there are four such states. With
the initial state $\ket{0}\otimes(\ket{H}-\ket{V})/\sqrt{2}$, the
overlap increases from $0$ to $0.828$ with $\phi\in(0^o,135^0)$ and
decrease to $0.8$ with $\phi\in(135^o,180^o)$. The localization
effect occurs in the range $\phi\in\left[45^o,180^o\right]$ and
becomes most notable with $\phi=135^o$.

\begin{figure}
\includegraphics[width=4.2cm]{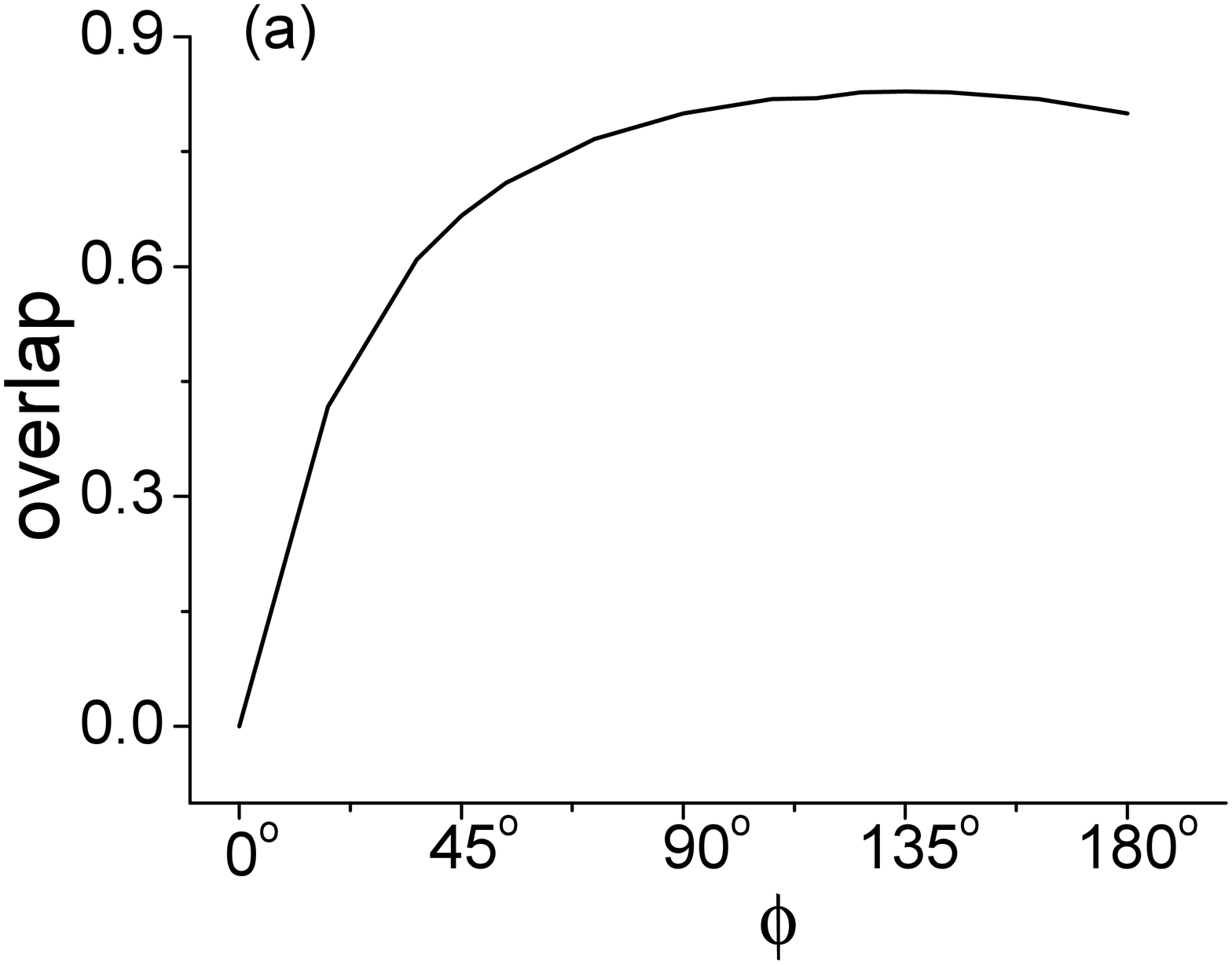}
\includegraphics[width=4.2cm]{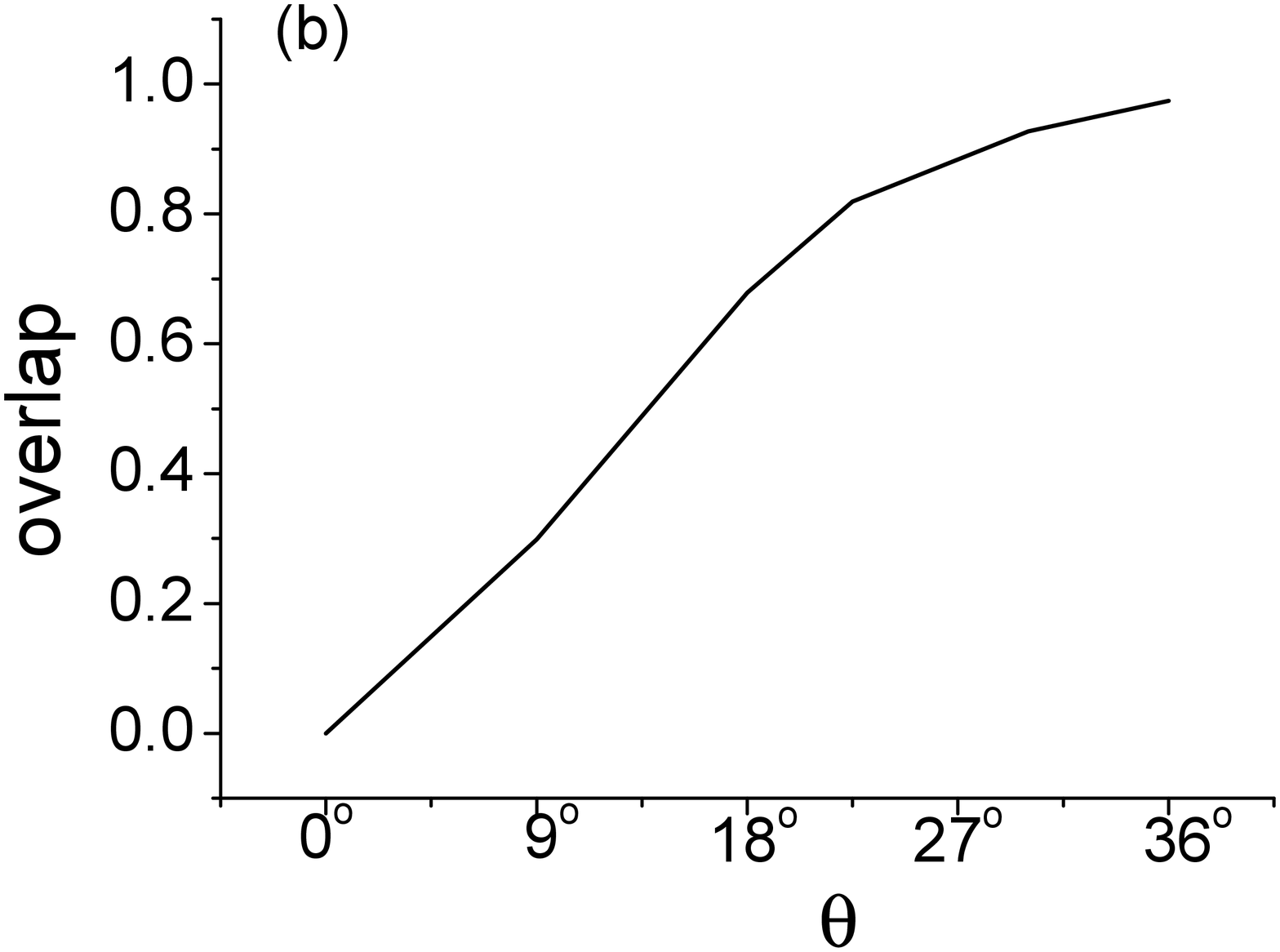}
\caption{(a) Overlap between the localized eigenstates of the
unitary operation $U$ of the Hadamard QW with SPPD in $x=0$ and the
initial state of the walker+coin system as a function of the phase
$\phi$. (b) Overlap between the localized eigenstates of $U$ of the
QW with SPPD $\phi=180^o$ in $x=0$ and the initial state as a
function of the coin bias $\theta$.}\label{Fig4}
\end{figure}

\begin{figure}
\includegraphics[width=4.2cm]{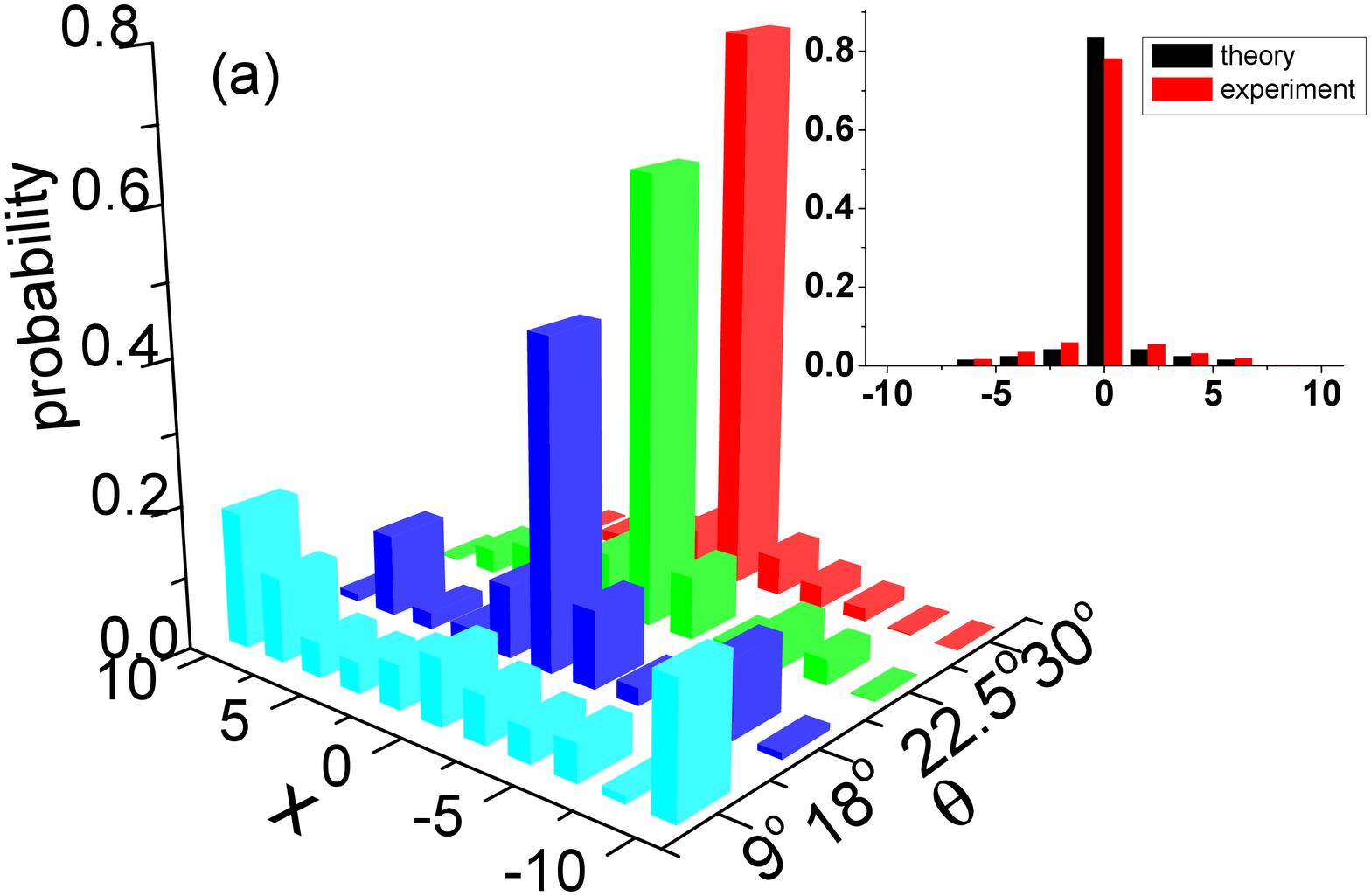}
\includegraphics[width=4.2cm]{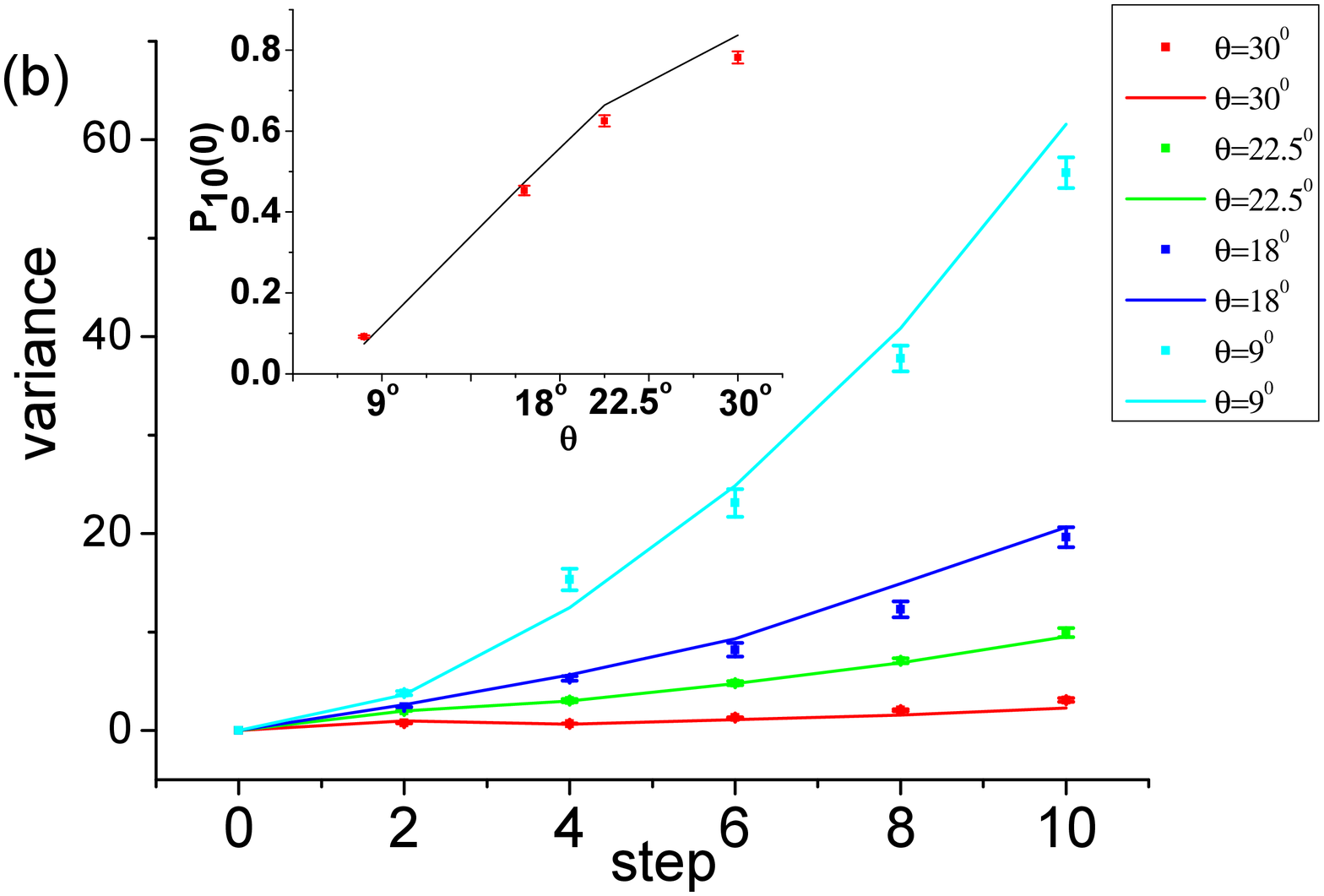}
\caption{(Color online.) (a) Experimental data of probability
distributions of the $10$-step QW with SPPD $\phi=180^o$ in the
original position $x=0$, antisymmetric initial coin state and
various coin bias $\theta$. The inset shows the probability
distribution of the $10$th-step QW with the coin bias $\theta=30^o$.
Red and black bars show experimental data and theoretical
predictions respectively. (b) Measured position variance of the
localized QW with antisymmetric initial coin state for $1$ to $10$
steps, with respective theoretical simulation (solid lines). The
inset shows the measured probabilities of the walker returning to
the original position $P_{10}(0)$ at the $10$th step with the
antisymmetric initial coin state as a function of the coin bias
$\theta$. Some of the statistical error bars are smaller than the
symbol size.}\label{Fig5}
\end{figure}

\begin{figure}
\includegraphics[width=4.2cm]{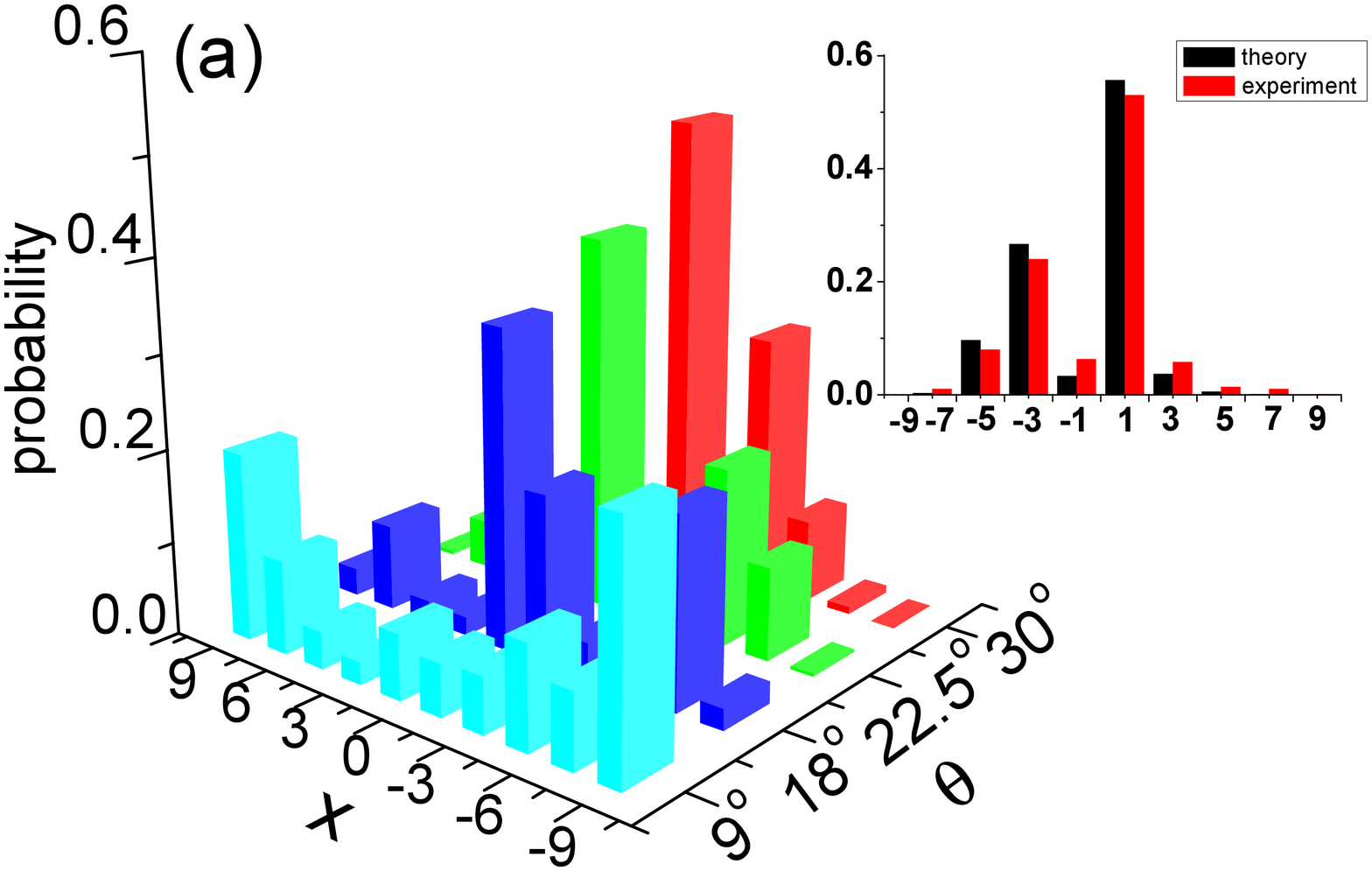}
\includegraphics[width=4.2cm]{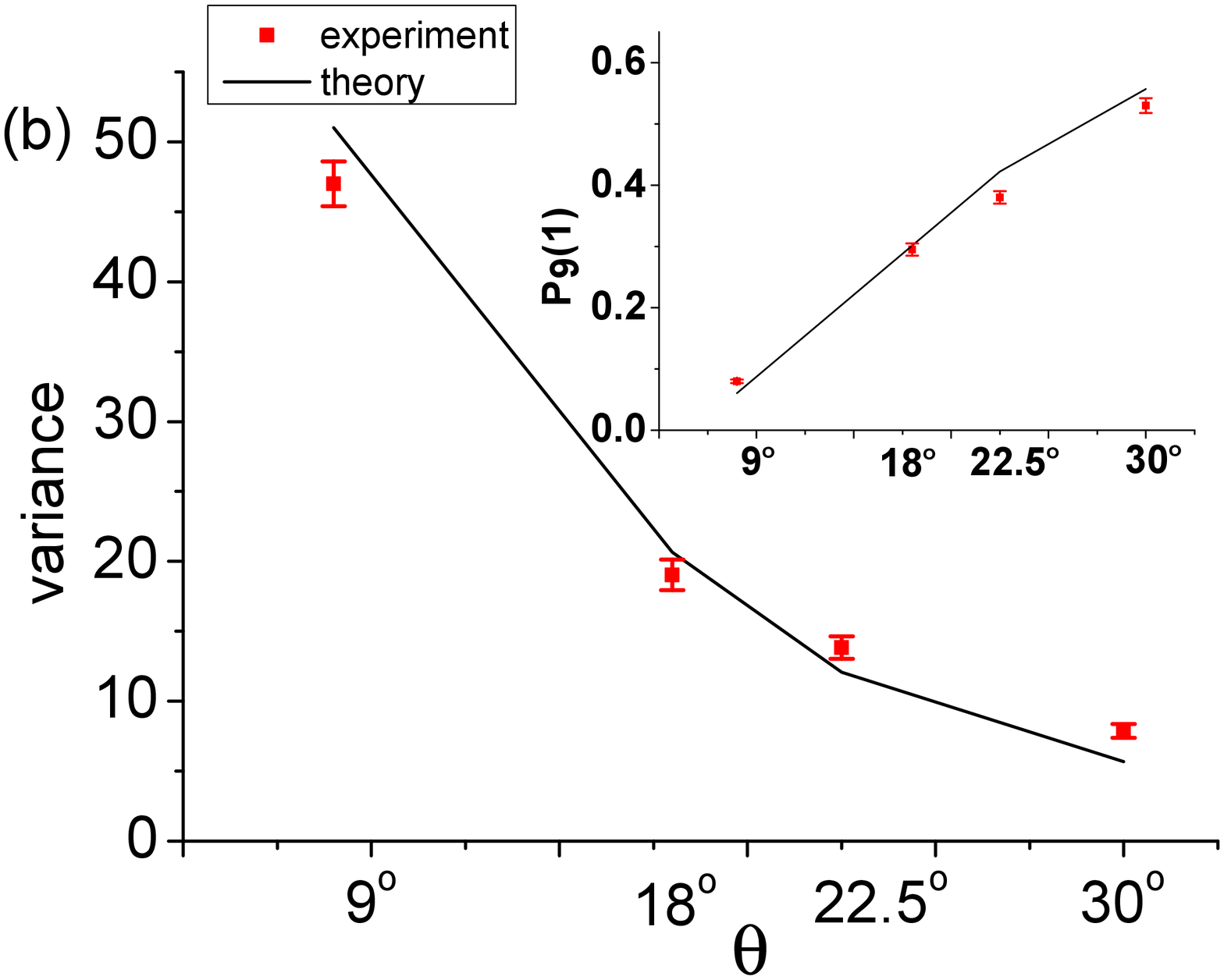}
\caption{(Color online.) (a) Experimental data of probability
distributions of the $9$-step QW with antisymmetric initial coin
state, SPPD $\phi=180^o$ in the position $x=1$ and various coin bias
$\theta$. The inset shows the probability distribution of the
$9$th-step QW with the coin flipping $C(30^o)$. Red and black bars
show experimental data and theoretical predictions respectively. (b)
Measured position variance of the $9$-step QW v.s. the coin bias
$\theta$, with respect to the theoretical predictions (solid line).
The inset shows the measured probabilities of the walker returning
to the $x=1$ position $P_9(1)$.}\label{Fig6}
\end{figure}

The next experimental result shown in Fig.~5 highlights the
flexibility of our implementation with respect to the easy
adjustability of the coin bias. Now we study the impact of the
different coin biases on the localized QWs. Fig.~5a shows the
measured position distributions for the $10$-step localized QW with
antisymmetric initial coin state, SPPD $\phi=180^o$  in the original
position $x=0$ and different coin biases $\theta=9^o, 18^o, 22.5^o,
30^o$ realized via different HWP settings. The walker is trapped in
the original position with the proper choice of $\phi$, which can
also be observed from the position variances and recurrence
probabilities in Fig.~5b. With the angle of HWP $\theta$ increasing,
the mode of behaviour of the walker+coin system is transmitted from
the diabatic (diabatic transition probability
$D=\cos^22\theta\sim1$) to adiabatic limit ($D\ll 1$). Harmin
predicted that suppression of diffusion occurs in the form of almost
perfect recurrences of the initial level population in the adiabatic
limit, while in the diabatic limit, the recurrence will
vanish~\cite{Harmin}. Our experiment result agrees with the
theoretic predictions and shows that with the HWP angle of
$\theta=9^o$ and $D=0.905$, in the diabatic limit the walker spreads
widely and no recurrence occurs. Whereas, with the HWP angle
increasing to $\theta=30^o$ ($D=0.095$), the mode of the behaviour
of the system is transmitted to the adiabatic limit, the diffusion
is more suppressed and the recurrence probability increases. Thus
the localization effect becomes more obvious in the adiabatic limit.
This result can also be explained by the dependence of the overlap
between the localized eigenstates and the initial state of the
walker+coin system on the coin bias shown in Fig.~4b. With $\theta$
increasing from $0^o$ to $30^o$, the overlap increases monotonically
from $0$ to $0.974$, which suggests the localization effect becomes
more notable.

Compared to the previous experimental results which only simulated
Anderson localization of the QW by trapping the walker in the
original position, we experimentally localize the single-photons in
different positions, for example in $x=1$. By inserting the PSs with
proper effective thickness in the spatial mode $x=1$ at each odd
step, we can realize a localized QW with SPPD $\phi=180^o$ in $x=1$.
In Fig.~6, the walker appears in $x=1$ with highest probability
after $9$ steps. With different coin settings, the localization
effect becomes more notable when the system is transmitted from
diabatic to adiabatic limit. The tendency of the position variance
and the probability of the walker back to $x=1$ depends on the coin
bias $\theta$ in the same manner compared to the case of the
localization in the original position.

The performance of our setup is limited only by imperfections of the
optical components such as nonplanar optical surfaces and the
coherence length of single photons, resulting in errors and
decoherence. The most significant source of systematic errors in our
setup is the imperfect coherence visibility of the BD
interferometer. A limitation for the maximal step number is given by
the size of the clear aperture of BDs. For example, for $10$-step
QW, the effective diameter of the clear aperture of the BD with beam
separation $3$mm needs to be larger than $30$mm. However this
problem is not intrinsic to this implementation, since the BDs with
large enough clear aperture and strictly planar surface can realize
the large-step QW.

In summary, we implement a stable and efficient way to realize QWs
embedded in a broader framework and show the phase defects can
influence the evolution of wavepackets. The QW with SPPD has the
single-photons localized in the certain position. Our experiment
benefits from the high stability and full control of both coin and
walker at each step and in each given position. The versatility of
our setup allows for extensions, such as the realization of
multi-particle QWs, in which richer choices of coin flipping and
defects would help us to study the topology of arbitrary graphs and
develop the applications such as quantum state transfer and energy
transportation problems. The localization can also be used to filter
and to trap particles, which would find applications in quantum
algorithms and quantum state engineering.

\acknowledgements We would like to thank Kai Sun, Xiaoye Xu, Jingshi
Xu, Chuanfeng Li and Barry C. Sanders for stimulating discussions.
This work has been supported by NSFC under 11174052, the Open Fund
from the SKLPS of ECNU and 973 Program under 2011CB921203.


\begin{references}

\bibitem{ADZ93} Y. Aharonov, L. Davidovich, and N. Zagury, Phys. Rev. A {\bf 48},
1687 (1993).
\bibitem{Ambainis} A. Ambainis, International Journal of Quantum Information, {\bf 1}, 507-518
(2003).
\bibitem{Spielman} A. M. Childs, R. Cleve, E. Deotto, E. Farhi, S. Gutmann,
and D. A. Spielman, Proc. 35th ACM Symposium on Theory of Computing
(STOC 2003), pp. 59-68.
\bibitem{Whaley02} N. Shenvi, J. Kempe, and K. B. Whaley, Phys. Rev. A {\bf 67}, 052307 (2003).
\bibitem{Kempe03} J. Kempe, Contemporary Physics {\bf44}, 307
(2003).
\bibitem{Childs09} A. M. Childs, Phys. Rev. Lett. {\bf102}, 180501
(2009).
\bibitem{Childs13} A. M. Childs, D. Gosset, Z. Webb, Science {\bf339}, 791-794
(2013).
\bibitem{Lovett10} N. B. Lovett, S. Cooper, M. Everitt, M. Trevers,
V. Kendon, Phys. Rev. A. {\bf81}, 042330 (2010).

\bibitem{KRBE10} T. Kitagawa, M. S. Rudner, E. Berg, and E. Demler, Phys. Rev. A {\bf 82}, 033429 (2010).
\bibitem{OPR06} A. C. Oliveira, R. Portugal, and R. Donangelo, Phys. Rev. A {\bf 74}, 012312 (2006).
\bibitem{HSW10} S. Hoyer, M. Sarovar, and K. B. Whaley, New J. Phys. {\bf12},
065041 (2010).
\bibitem{A58} P. W. Anderson, Phys. Rev. {\bf109}, 1492 (1958).
\bibitem{BCA03} T. A. Brun, H. A. Carteret, and A. Ambainis, Phys. Rev. Lett. {\bf91}, 130602
(2003).
\bibitem{W12} A. W\'{o}jcik, T. {\L}uczak, P. Kurzy\'{n}ski, A. Grudka, T. Gdala, and M. Bednarska-Bzdega, Phys. Rev. A {\bf 85}, 012329
(2012).
\bibitem{ES11} E. Segawa, arXiv: 1112.4982.
\bibitem{YKE08} Y. Yin, D. E. Katsanos, and S. N. Evangelou, Phys. Rev. A {\bf 77}, 022302
(2008).
\bibitem{C13} A. Crespi, R. Osellame, R. Ramponi, V. Giovannetti, R. Fazio, L. Sansoni,
F. De Nicola, F. Sciarrino, and P. Mataloni, Nature Photonics
{\bf7}, 322-328 (2013).
\bibitem{SS11} A. Schreiber, K. N. Cassemiro, V. Poto\u{c}ek, A. G\'{a}bris, I. Jex, and Ch. Silberhorn, Phys. Rev. Lett. {\bf106}, 180403
(2011).
\bibitem{WM04} A. W\'{o}jcik, T. {\L}uczak, P. Kurzy\'{n}ski, A. Grudka, and M. Bednarska, Phys. Rev. Lett. {\bf93}, 180601
(2004).
\bibitem{BB04} O. Byerschaper, and K. Burnett, arXiv: quant-ph/0406039.
\bibitem{B06} M. C. Ba\~{n}uls, C. Navarrete, A. P\'{e}rez,
E. Rold\'{a}n, and J. C. Soriano, Phys. Rev. A {\bf 73}, 062304
(2006).
\bibitem{Xue13} P. Xue, H. Qin, B. Tang, and B. C. Sanders, arXiv:
1312.0123.
\bibitem{B99} D. Bouwmeester, I. Marzoli, G. P. Karman, W. Schleich,
and J. P. Woerdman, Phys. Rev. A {\bf 61}, 013410 (1999).
\bibitem{DH03} J. Du, H. Li, X. Xu, M. Shi, J. Wu, X. Zhou, and R. Han, Phys. Rev. A {\bf 67}, 042316
(2003).
\bibitem{ZR10} F. Z\"{a}hringer, G. Kirchmair, R. Gerritsma, E. Solano, R. Blatt, and C. F. Roos, Phys. Rev. Lett. {\bf104}, 100503 (2010).
\bibitem{SM09} H. Schmitz, R. Matjeschk, C. Schneider, J. Glueckert,
M. Enderlein, T. Huber, and T. Schaetz, Phys. Rev. Lett. {\bf103},
090504 (2009).
\bibitem{KF09} M. Karski, L. Forster, J. M. Choi, A. Steffen, W. Alt,
D. Meschede, and A. Widera, Science {\bf325}, 174 (2009).
\bibitem{CG06} R. Cote, A. Russell, E. E. Eyler, and P. L. Gould, New J. Phys. {\bf8}, 156 (2006).
\bibitem{D05} B. Do, M. L. Stohler, S. Balasubramanian, D. S.
Elliott, C. Eash, E. Fischbach, M. A. Fischbach, A. Mills, and B.
Zwickl, J. Opt. Soc. Am. B {\bf 22}, 499 (2005).
\bibitem{Z07} P. Zhang, X. F. Ren, X. B. Zou, B. H. Liu, Y. F.
Huang, and G. C. Guo, Phys. Rev. A {\bf 75}, 052310 (2007).
\bibitem{PO10} A. Peruzzo, M. Lobino, J. C. F. Matthews, N. Matsuda, A. Politi,
K. Poulios, X. Zhou, Y. Lahini, N. Ismail, K. Worhoff, Y. Bromberg,
Y. Silberberg, M. G. Thompson, and J. L. O'Brien, Science {\bf329},
1500 (2010).
\bibitem{PS08} H. B. Perets, Y. Lahini, F. Pozzi, M. Sorel, R. Morandotti, and Y. Silberberg, Phys. Rev. Lett. {\bf100}, 170506
(2008).
\bibitem{SS12} A. Schreiber, A. G\'{a}bris, P. P. Rohde, K. Laiho, M. \u{S}tefa\u{n}\'{a}k,
V. Poto\u{c}ek, C. Hamilton, I. Jex, and C. Silberhorn, Science
{\bf336}, 55 (2012).
\bibitem{BF+10} M. A. Broome, A. Fedrizzi, B. P. Lanyon, I. Kassal,
A. Aspuru-Guzik, and A. G. White, Phys. Rev. Lett. {\bf 104}, 153602
(2010).
\bibitem{S10} A. Schreiber, K. N. Cassemiro, V. Poto\u{c}ek, A. G\'{a}bris, P. J. Mosley, E. Andersson, I. Jex, and Ch. Silberhorn, Phys. Rev.
Lett. {\bf 104}, 050502 (2010).
\bibitem{SSV+12}  L. Sansoni, F. Sciarrino, G. Vallone, P. Mataloni, A. Crespi,
R. Ramponi, and R. Osellame, Phys. Rev. Lett. {\bf 108}, 010502
(2012).
\bibitem{Harmin} D. A. Harmin, Phys. Rev. Lett. {\bf 56}, 232
(1997).


\end{references}
\end{document}